\documentclass[journal,twoside]{IEEEtran}
\ifCLASSINFOpdf
\else
\fi

\usepackage{cite}
\usepackage{amsmath,amssymb,amsfonts}
\usepackage{algorithmic}
\usepackage{graphicx}
\usepackage{textcomp}

\usepackage[utf8]{inputenc}
\usepackage{physics}
\usepackage{lineno}
\usepackage{cite}
\usepackage{enumerate,subfigure}
\usepackage{latexsym}
\usepackage{framed}
\usepackage{hyperref}
\usepackage{cite}
\usepackage{afterpage}
\usepackage{comment}
\usepackage{multirow}

\usepackage[dvipsnames]{xcolor}

\DeclareMathOperator{\lamTV}{\lambda_{\mbox{\tiny TV}}}
\DeclareMathOperator{\lamLM}{\lambda_{\mbox{\tiny LM}}}
\DeclareMathOperator{\REsig}{RE_\sigma^{\ell_1}}
\DeclareMathOperator{\REv}{RE_{V}^{\ell_2}}
\DeclareMathOperator{\MSEsig}{MSE_\sigma}

\begin{document}

%
\title{Graph Convolutional Networks for Model-Based Learning in Nonlinear Inverse Problems}
%
%
%

\author{William Herzberg, Daniel B. Rowe, Andreas Hauptmann,~\IEEEmembership{Member, IEEE}, and Sarah J. Hamilton
\thanks{W. Herzberg, D. Rowe, and S.J. Hamilton are with the Department of Mathematical and Statistical Sciences; Marquette University, Milwaukee, WI 53233 USA,  \texttt{(william.herzberg@mu.edu)},  \texttt{(daniel.rowe@mu.edu)}, \texttt{(sarah.hamilton@mu.edu)}}%
\thanks{A. Hauptmann is with the Research Unit of Mathematical Sciences; University of Oulu, Oulu, Finland and with the Department of Computer Science; University College London, London, United Kingdom, {\tt (e-mail: andreas.hauptmann@oulu.fi)}}%
\thanks{Manuscript submitted: July 2, 2021. 
This work was supported in part by National Institute Of Biomedical Imaging And Bioengineering of the National Institutes of Health under Award Number R21EB028064. The work of AH is supported by the Academy of Finland project numbers: 336796, 338408.}}

%
%

\markboth{}
{Herzberg \MakeLowercase{\textit{et al.}}: Graph Convolutional Networks for Model-Based Learning}
%



\maketitle


\begin{abstract}
The majority of model-based learned image reconstruction methods in medical imaging have been limited to uniform domains, such as pixelated images. If the underlying model is solved on nonuniform meshes, arising from a finite element method typical for nonlinear inverse problems, interpolation and embeddings are needed. To overcome this, we present a flexible framework to extend model-based learning directly to nonuniform meshes, by interpreting the mesh as a graph and formulating our network architectures using graph convolutional neural networks. This gives rise to the proposed iterative Graph Convolutional Newton-type Method (GCNM), which includes the forward model in the solution of the inverse problem, while all updates are directly computed by the network on the problem specific mesh.  We present results for Electrical Impedance Tomography, a severely ill-posed nonlinear inverse problem that is frequently solved via optimization-based methods, where the forward problem is solved by finite element methods. Results for absolute EIT imaging are compared to standard iterative methods as well as a graph residual network. We show that the GCNM has strong generalizability to different domain shapes and meshes, out of distribution data as well as experimental data, from purely simulated training data and without transfer training. 
\end{abstract}


%
\IEEEpeerreviewmaketitle

\section{Introduction}\label{Sec: Introduction}
\IEEEPARstart{M}{any} tomographic image reconstruction tasks fit under the umbrella of \textit{inverse problems} as they seek to recover an image $x$ from indirect measurements $y$. Further, these are typically ill-posed which means that finding a unique solution in a stable manner ranges from difficult to impossible without the use of prior knowledge about the problem. With the rise of deep learning, classical reconstruction techniques have been paired with, or replaced by, learned methods that can increase stability and promote uniqueness by enforcing strong data-driven priors. Specifically, supervised learning is still the most common approach for medical image reconstruction tasks, where a data set of known ground truth images and measurement input pairs are used to train a network that is later used to compute predictions from new inputs. When the forward operator $\mathcal{T}$ for the imaging task is known, a set of training data can be simulated, which is especially useful when a large number of experimental training samples are not available.

Learned image reconstruction can be broken down further into three main categories\cite{arridge2019solving,Ongie2020}. The first, and most common, uses an analytically known inversion operator $\mathcal{T}^{\dagger}$ to obtain an approximate reconstruction that can suffer from noise or other artefacts such as distortions. Then, a neural network $\Lambda_{\theta}$ with parameters $\theta$ can be used, with the approximate solution as the input, to improve this initial reconstruction and obtain a cleaned version as, \cite{kang2017deep,jin2017deep},   
\[ \hat{x}= \Lambda_{\theta} \left( \mathcal{T}^{\dagger} y \right).\]  
The second category is commonly referred to as iterative \textit{model-based} techniques (or unrolled methods) because information about the known forward model $\mathcal{T}$ is intertwined with learned components in an iterative manner. At each such iterative step, a neural network $\Lambda_{\theta_{k}}$ computes an updated reconstruction that may use a combination of the current iterate, measurements, and forward model, as inputs \cite{putzky2017recurrent,adler2017solving}: 
\[x_{k+1} = \Lambda_{\theta_{k}} \left( x_{k}, y, \mathcal{T} \right).\]  
A third category might use a neural network to compute the image directly from the measurements \cite{zhu2018image}. This approach is far less common in imaging tasks, due to training instabilities, the need for large data, and limited generalization capabilities with respect to changes in the measurement setup \cite{baguer2020computed}.

The majority of learned image reconstruction tasks use convolutional neural networks (CNNs), which are especially favorable for imaging applications due to their translational invariance and capability to leverage local dependencies and structures. Applications in medical imaging have been limited to pixel/voxel grids, but for many nonlinear inverse problems, the forward model is solved using the finite element method (FEM) which requires discretizing the domain onto special meshes. These meshes often have triangular elements and are very irregular, unlike the pixel grids needed for the application of a CNN. To incorporate CNNs into imaging tasks where the image is initially defined over such meshes, one needs to perform an interpolation step, or equivalent, to convert mesh data to pixel-grid data and embed it into a rectangular domain. Consequently, it would be more natural to perform the learned image reconstruction directly on the domain and geometry defined by the FEM mesh. 

Hence, we propose considering the data defined over a FEM mesh as graph data so that graph convolutional networks (GCNs) can be used as an alternative to the traditional CNN in a model-based learned approach for solving nonlinear inverse problems. In the simplest form, graph data is composed of nodes connected by edges. In this work, non-directional, unweighted, homogeneous graphs are considered. GCNs were designed specifically for graph data and are intended to leverage many of the same benefits as traditional CNNs; shared weights, translational invariance, and localization \cite{KipfWelling2017}. 
Here, we apply GCNs to FEM mesh data in the context of learned image reconstruction, opening a new avenue to work directly on the problem specific meshes.

The proposed method, dubbed the Graph Convolutional Newton-type Method (GCNM), utilizes the model-based learning approach by incorporating GCNs into traditional iterative methods for solving nonlinear inverse problems, namely Newton-type methods, which use FEM meshes to solve the forward problem. This enables us to leverage the advantages offered by convolutional networks, while being able to operate directly on the FEM meshes used in the forward problem. This eliminates the need to convert between meshes and pixel grids, and work more naturally on the underlying domain geometry. Furthermore, using the forward model at each iteration reaps many of the advantages seen in other model-based methods. 

In this work we show that the GCNM has strong fitting abilities, requires fewer iterations than classical optimization-based methods, and can generalize well to new domain shapes, FEM meshes, and noise patterns without the need for transfer training. We will compare our approach to classical optimization based methods, as well as a post-processing ResNet with graph convolutional layers (GResNet), which will showcase the importance of repeated incorporation of model information at each iteration.

In the following, we first motivate and introduce the novel GCNM as well as the post-processing GResNet used for comparison.  Next, Section~\ref{Sec: Electrical Impedance Tomography}, reviews the mathematical problem of electrical impedance tomography (EIT), a highly ill-posed nonlinear inverse problem, which will serve as a challenging case study for the proposed GCNM.  Section~\ref{Sec: Methods} describes the examples considered, training data used, and evaluation metrics that will be used to assess the results.  Results for simulated and experimental data are presented in Section~\ref{Sec: Results}, and conclusions are drawn in Section~\ref{Sec: Conclusion}.

\section{Graph Convolutional Newton-type Method} \label{Sec: Graph Convolutional Newton's Method}
%
%
\begin{figure*}[h!]
    \centering
    \begin{picture}(500,230) 
    
    \put(2.5,8){\includegraphics[width=500pt]{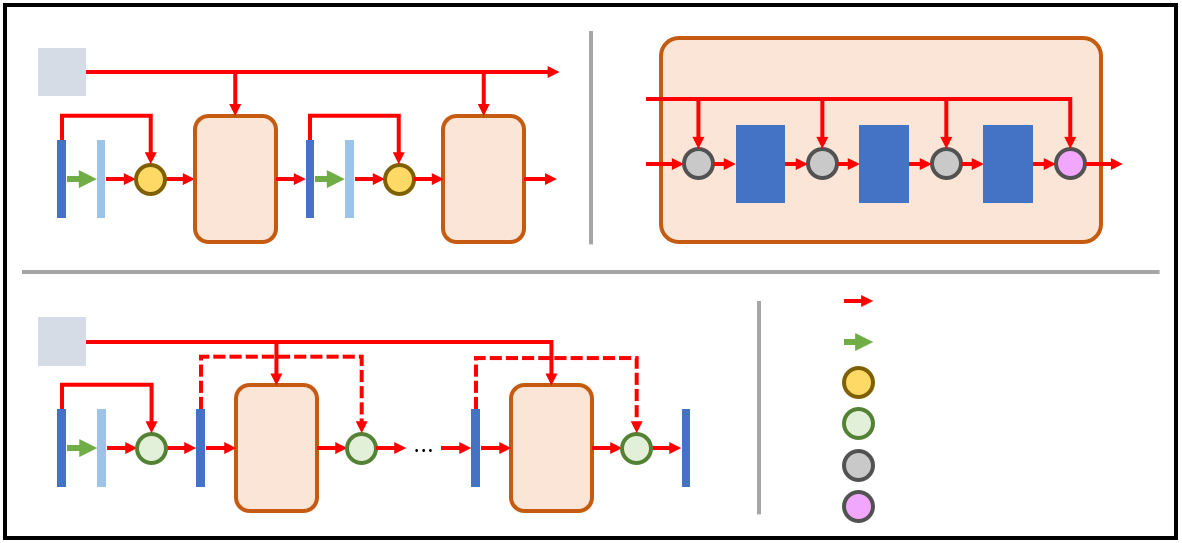}}
    
    \put(25,223){ \sc Graph Convolutional Newton-type Method}
    \put( 26,203){\footnotesize \sc $\mathcal{A}$}
    \put( 95,159){\footnotesize \sc $\Lambda_{\theta_{0}}$}
    \put(200,159){\footnotesize \sc $\Lambda_{\theta_{1}}$}
    \put( 24,133){\footnotesize \sc $x_{0}$}
    \put( 40,133){\footnotesize \sc $\delta x_{0}$}
    \put(129,133){\footnotesize \sc $x_{1}$}
    \put(145,133){\footnotesize \sc $\delta x_{1}$}
    
    \put(343,208){\sc Block k, $\Lambda_{\theta_{k}}$}
    \put(264,192){\footnotesize \sc $\mathcal{A}$}
    \put(256,165){\footnotesize \sc $H^{(0)}$}
    \put(315,140){\footnotesize \sc $H^{(1)}$}
    \put(368,140){\footnotesize \sc $H^{(2)}$}
    \put(420,140){\footnotesize \sc $H^{(3)}$}
    \put(478,165){\footnotesize \sc $H^{(4)}$}
    
    \put(55,110){ \sc Graph Convolutional Residual Network}
    \put( 26,90){\footnotesize \sc $\mathcal{A}$}
    \put(112, 45){\footnotesize \sc $\Lambda_{\theta_{0}}$}
    \put(222, 45){\footnotesize \sc $\Lambda_{\theta_{k_{max}}}$}
    \put( 24, 20){\footnotesize \sc $x_{0}$}
    \put( 40, 20){\footnotesize \sc $\delta x_{0}$}
    \put( 83, 20){\footnotesize \sc $x_{1}$}
    \put(287, 20){\footnotesize \sc $x_{rec}$}
    
    \put(380,107){\sc Input/Output}
    \put(380, 90){\sc Compute $\delta x$}
    \put(380, 72){\sc Concatenation}
    \put(380, 55){\sc Addition}
    \put(380, 37){\sc ReLU(GraphConv)}
    \put(380, 19){\sc GraphConv}
    
    \end{picture}
    \caption{\label{Fig: network diagram} (Top Left) The first two iterations of the GCNM are shown. (Bottom Left) The GResNet which takes the first iteration of a classical Newton method as input to the first block and a total of $k_{max}$ blocks is shown. (Top Right) One block of a network is shown. The same block structure is used in both GCNM and the GResNet. In GCNM, the input $H^{(0)}$ is the concatenation of $x_{k}$ and $\delta x_{k}$ and the output $H^{(0)}$ is $x_{k+1}$. For the GResNet, $H^{(0)}$ is $x_{1}$ for the first block and the sum of the previous block's input and output for the remaining blocks.}
\end{figure*}
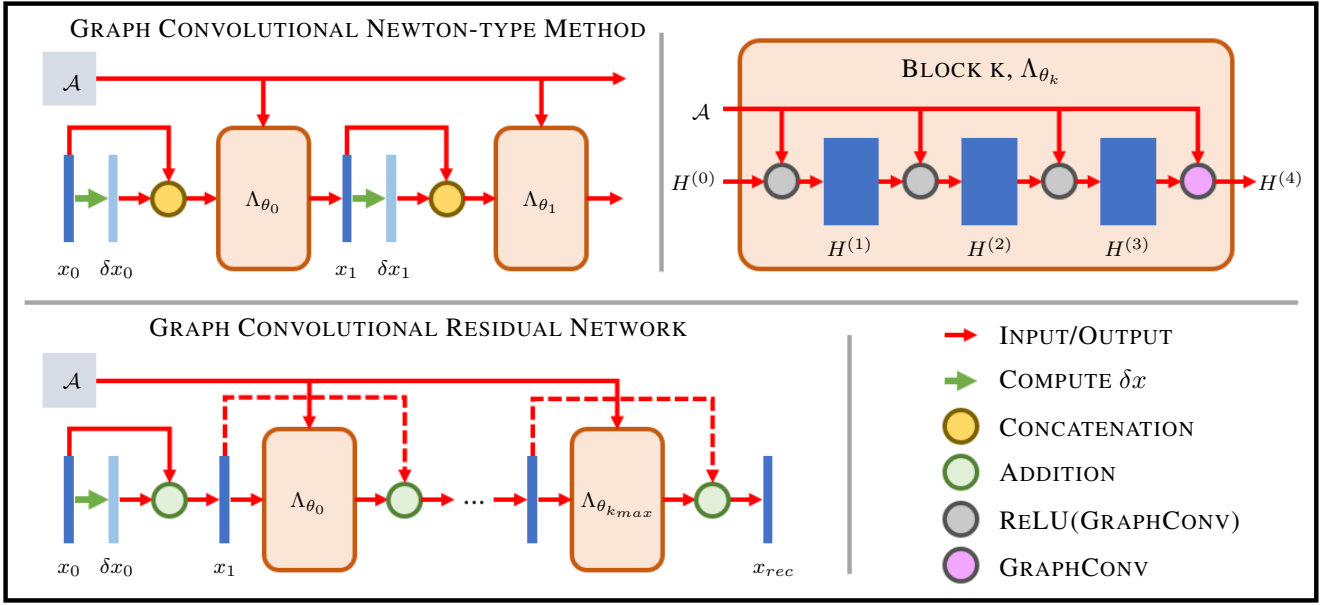    

The proposed Graph Convolutional Newton-type Method (GCNM) is a new model-based image reconstruction technique that operates on FEM meshes and builds on iterative second order (Newton-type) methods but changes how the current iterate $x_{k}$ and its update $\delta x_{k}$ are combined. Classic methods obtain the next iterate as $x_{k+1} = x_{k} + \delta x_{k}$, where $\delta x_k$ is computed according to a chosen optimization method such as Gauss-Newton (GN) or Levenberg-Marquardt (LM). The GCNM, by contrast, computes the next iterate using a trained network $\Lambda_{\theta_k}$ and the adjacency matrix $\mathcal{A}$, for the graph over which $x_{k}$ and $\delta x_{k}$ are defined, via
\begin{equation}\label{Eq: network k}
    x_{k+1} = \Lambda_{\theta_{k}} \left( \left[ x_{k}, \delta x_{k}  \right], \mathcal{A} \right),
\end{equation}
where $\delta x$ is the traditional update (e.g. GN, LM). The two terms $x_{k}$ and $\delta x_{k}$ are concatenated and then used as input to a graph convolutional (GCN) block. The output of the GCN block provides the next iterate $x_{k+1}$, see Figure~\ref{Fig: network diagram}. In this work, the structure of the blocks $\Lambda_{\theta_k}$ are the same at each iteration but each block will have its own unique set of trainable parameters $\theta_{k}$. In this study we chose the LM updates for the GCNM, to not enforce strong data specific priors and rather expect the networks to learn the features from the training set. We note that the proposed framework extends to other updates arising from GN methods and one could include stronger priors, such as total variation (TV), if desired.

As with other learned model-based methods \cite{putzky2017recurrent,adler2017solving,hauptmann2018model,hauptmann2020multi,Lin2020} there are two options for training the network: train the entire system of $k_{max}$ blocks end-to-end, or train each block sequentially. In our case, an end-to-end training is not practical for two main reasons. First, to update the network parameters we would need to perform back-propagation through the updates $\delta x_k$, computed by evaluating the model equations of the underlying problem given by a FEM solver, which is not practical. This leads directly to the second problem, that evaluating the model equations is time consuming and would lead to extensive training times. Thus, we follow here the sequential approach of training each block separately \cite{hauptmann2018model}. 
Given a training set of true $x_{true, (i)}$ and current iterate $x_{k,(i)}$ pairs for $i=1,\dots,M$, this leads then to a loss function requiring iterative-wise optimality
\begin{equation}\label{Eq: loss fn}
    \mathrm{Loss}(\theta_{k}) = \frac{1}{M} \sum_{i=1}^{M} ( \Lambda_{\theta_{k}} \left( \left[ x_{k}, \delta x_{k}  \right], \mathcal{A} \right)_{(i)} -x_{true, (i)} )^{2}. 
\end{equation}

In order to consider a quantity $x$ and update $\delta x$, defined on a FEM mesh with $M$ elements, as graph data, two pieces are needed: a feature matrix and an adjacency matrix. The feature matrix $H \in \mathbb{R}^{M \times f}$ has one row for each graph node and one column for each feature defined over the nodes. The adjacency matrix $\mathcal{A} \in \mathbb{R}^{M \times M}$ is sparse and describes how the graph nodes are connected. Only the entries $\mathcal{A}_{ij}=1$ are nonzero where graph nodes $i$ and $j$ are connected. Here, we consider each mesh element as a graph node and two elements are connected if they share at least one mesh node.  Alternatively, if the solution is defined at the FEM nodes, those would be the natural graph nodes to use.  

%
\subsection{Graph Convolutions}\label{SubSec: GraphConv}
%
Kipf and Welling \cite{KipfWelling2017} proposed a convolutional layer for neural networks that operates on graph data and is analogous to the convolutional layer used on pixel grid data, given input $H^{(i)}$ the layer is defined as: 
\begin{equation}\label{Eq: GCN propagation}
    H^{(i+1)} = g 
    \left( 
    \Tilde{D}^{-\frac{1}{2}} \Tilde{\mathcal{A}} \Tilde{D}^{-\frac{1}{2}} H^{(i)} W^{(i)} \right).
\end{equation}
Self loops, or connections from graph node $i$ to itself, are included in the adjacency matrix by $\Tilde{\mathcal{A}} = \mathcal{A} + I$ and $\Tilde{D}^{-\frac{1}{2}}$ denotes the inverse of the square root of each element from $\Tilde{D}=\mathrm{diag}(\sum_{j}\Tilde{\mathcal{A}}_{ij})$. 
The weight matrix $W^{(i)} \in \mathbb{R}^{ f^{(i)} \times f^{(i+1)} }$ contains the trainable parameters for the layer with number of input $f^{(i)}$ and output $f^{(i+1)}$ features. Finally, $g\left(\cdot\right)$ denotes the nonlinear activation function.
If biases are desired, an extra column of ones $\vec{1}\in\mathbb{R}^{M}$ is concatenated to the 
input feature matrix and an extra row of trainable parameters $b^{i}\in\mathbb{R}^{f(i+1)}$ is concatenated to the weight matrix resulting in 
\begin{equation}\label{Eq: GCN propagation biases}
    H^{(i+1)} = g 
    \left( \Tilde{D}^{-\frac{1}{2}} \Tilde{\mathcal{A}} \Tilde{D}^{-\frac{1}{2}}
    \left[ H^{(i)} , \vec{1} \right]
    \left[ W^{(i)} ; b^{(i) T} \right]
    \right).
\end{equation}

In \eqref{Eq: GCN propagation} and \eqref{Eq: GCN propagation biases}, one can think of $\Tilde{D}^{-\frac{1}{2}} \Tilde{\mathcal{A}} \Tilde{D}^{-\frac{1}{2}} H^{(i)}$ as an aggregation of features within the neighborhood of each graph node. Then, multiplication with $W^{(i)}$ computes a node's output features as linear combinations of its aggregated input features. Stacking $X$ of these layers increases a node's receptive field to all of its $X$ neighborhood in the graph. 
Note that standard 2D convolutions learn linear combinations of neighbors' feature values for aggregating within a neighborhood while GCN layers only use a specific weighted average (described by $\Tilde{D}^{-\frac{1}{2}} \Tilde{\mathcal{A}} \Tilde{D}^{-\frac{1}{2}}$) for aggregating information that is not learned, but instead defined by the mesh.
%
%
\subsection{Network Structure}\label{SubSec: Network Structure}
%
At each iteration of the GCNM, a GCN block with the structure shown in Fig.~\ref{Fig: network diagram} is used. The input graph uses the $M$ elements from the mesh as nodes in the graph with an adjacency matrix also formed from the mesh. Each graph node has $f^{(0)}=2$ initial features such that $H^{(0)} = [x_{k}, \delta x_{k}]$. Then, 
three graph convolutional layers with 
ReLU activation functions
expand the feature dimension 
to 250 features before a final graph convolutional layer with linear activation is used to produce the output of the block.
The output is a graph with the same adjacency matrix as the input, but with only one feature $x_{k+1}$ defined over the nodes. Similar to \cite{adler2017solving,hauptmann2018model}, each block's network structure was chosen to be small and quite simple as compared to typically larger post-processing networks. Note, that in contrast to \cite{adler2017solving,hauptmann2018model}, we do not use a residual update. The above architecture has been found to work well for all test problems considered here; more specialised architectures could be considered for a particular problem but this is outside the scope of this current work.

One important aspect of the new GCNM is that it uses model information iteratively. That means, the output from one block $x_{k+1}$ is used to compute the next update $\delta x_{k+1}$, and then both are used as input in the next block $\Lambda_{\theta_{k+1}}$. At each iteration, new information from the forward problem is being introduced by $\delta x_{k}$ and a new GCN block acts on the inputs. This can be a strength when the original updates $\delta x_{k+1}$ converge toward the true solution. Alternatively, if the forward model is not accurate, the GCN can compensate and correct for the wrong components and extract useful information for the updates, acting as a learned model correction \cite{lunz2021learned,smyl2021learning}. We will see this correcting nature in the experiments (e.g. Fig.~\ref{fig:recons-outdist-dom}). 

Finally, when regularization parameters are used in the computation of the updates $\delta x_{k}$ that are used as input to the networks, the parameters can be tuned to new test data after training.  This will be demonstrated in the following experiments where different noise distributions, parameter ranges, and experimental data are considered.

\section{Case Study: Electrical Impedance Tomography}\label{Sec: Electrical Impedance Tomography}

We chose Electrical Impedance Tomography (EIT) as a case study for the proposed GCNM due to its nonlinear nature and severe ill-posedness.  EIT is an imaging modality that uses electrodes attached to the surface of a domain to inject current and measure the resulting electrical potential. The electrical measurements are used to recover the conductivity distribution $\sigma$ inside the domain.  The physical problem in EIT can be modeled via the {\it conductivity equation}~\cite{Calderon1980}
\begin{equation}\label{Eq: eit main}
    \nabla \cdot \sigma \nabla u = 0 \quad \text{in $\Omega\subset\mathbb{R}^n$},
\end{equation}
where $u$ denotes the potential and $0<\sigma<\infty$. The recovery of the internal conductivity from surface electrical measurements is a severely ill-posed, nonlinear inverse problem which requires carefully designed numerical solvers robust to noise and modeling errors.

Most commonly, the reconstruction problem in EIT is formulated in a variational setting and solved by iterative methods that minimize the error between measured voltages $V$ and simulated voltages $U(\sigma)$ corresponding to a guess conductivity $\sigma$. Additional regularization is needed due to the ill-posedness and instability of the reconstruction problem; popular approaches include the Levenberg-Marquardt (LM) algorithm \cite{li2020levenbergmarquardt}, Tikhonov regularization \cite{Lionheart2004}, and Total Variation (TV) regularization \cite{Borsic2010}. These methods frequently suffer from high sensitivity to modeling errors and, if not compensated for, are limited in practice to {\it time-difference} EIT imaging which recovers the change in conductivity $\Delta\sigma$ relative to a reference data set/frame. The most common applications of time-difference EIT focus on monitoring heart and lung function of hospitalized patients.  {\it Absolute}, also called {\it static}, EIT imaging recovers the conductivity at the time of measurement from a single frame $V$ of EIT data. While in time-difference imaging, some of the modeling errors can cancel out, in absolute imaging they do not and hence require a correction \cite{nissinen2011compensation,dong2015estimation}. Developing fast robust image reconstruction algorithms for absolute EIT imaging is important for applications such as breast cancer imaging, stroke classification, and nondestructive evaluation where a pre-injury set of measurements is unavailable; see \cite{Borcea2002a,Mueller2012} for a further literature review of applications.

Alternatively, direct (non-iterative) reconstruction methods such as the D-bar method \cite{Nachman1996,Hamilton2018_Robust,Mueller2020, Dodd2014}, and Calder\'on's method \cite{Calderon1980, Muller2017} show promise for fast robust absolute and time-difference EIT imaging. However, these methods often suffer from blurred reconstructions as a result of a low-pass filtering of the associated (non)linear Fourier data required by the reconstruction algorithms, and offer limited applicability in an iterative model-based learned reconstruction framework.

As with other computational imaging tasks, deep learning has been leveraged in many ways to improve EIT reconstruction quality while maintaining or reducing inference time. Several direct approaches in the category of post-processing based learned image reconstruction have been proposed, such as Deep D-bar methods \cite{Hamilton2018,Hamilton_2019} and the dominant-current approach \cite{wei2019dominant} where a CNN utilizing the popular U-net architecture \cite{ronneberger2015u} is trained and used to improve an initial reconstruction in the image space on a pixel grid. 
Alternatively, a neural network-based supervised descent method (SDM) described in \cite{Lin2020} falls into the \textit{model-based} category.  The inputs are residuals in the measurement space while the outputs are updates in the image space defined over a FEM mesh. Another model-based approach was presented in \cite{Smyl2021efficient}, where the authors propose a Quasi Newton method by learning updates of the Jacobians. The GCNM presented in Section~\ref{Sec: Graph Convolutional Newton's Method} also falls in the \textit{model-based} category, using GCNs with inputs and outputs in the image space defined over the problem specific FEM meshes used in the optimization method.

\subsection{Solving the Forward Problem in EIT}\label{SubSec: Forward Problem}
Given a domain $\Omega \subset \mathbb{R}^n$, the EIT forward problem is to determine the electrical potential $u$ at the boundary of the domain $\partial \Omega$ when current is applied and the conductivity distribution of the interior is known. The boundary conditions for the conductivity equation \eqref{Eq: eit main} are given by the complete electrode model \cite{Somersalo1992},
\begin{equation}\label{Eq: complete electrode model}
    \begin{cases}
        \int_{e_{\ell}} \sigma  \frac{ \partial u }{ \partial \hat{n} } dS = I_{\ell} , \quad \ell = 1, 2, ..., L ,\\
        \left. \sigma \frac{ \partial u }{ \partial \hat{n} } \right\vert_{\partial \Omega / \cup e_{\ell}} = 0 ,\\
        \left. ( u + z_{\ell} \sigma \frac{ \partial u }{ \partial \hat{n} } ) \right\vert_{e_{\ell}} = U_{\ell}, \quad \ell = 1, 2, ..., L, \\
        \sum\limits_{\ell=1}^{L} I_{\ell} = 0, \\
        \sum\limits_{\ell=1}^{L} U_{\ell} = 0, \\
    \end{cases}
\end{equation}
where $L$ is the number of electrodes and $e_{\ell}$ is the $\ell^{th}$ electrode; $z_{\ell}$, $I_{\ell}$, and $U_{\ell}$, are the contact impedance, current injected, and electric potential on the $\ell^{th}$ electrode, respectively; and $\hat{n}$ is the outward unit vector normal to the boundary. Following \cite{Kaipio2000,Vauhkonen1997}, the forward problem \eqref{Eq: eit main} and \eqref{Eq: complete electrode model} can be solved using FEM to determine the voltages $U(\sigma)$ on the electrodes for a given conductivity $\sigma$.  Here we consider the 2D ($n=2$) case. 

\subsection{The Inverse Problem for EIT}\label{SubSec: Inverse Problem}

Many iterative methods for EIT reconstruction have been proposed that begin as a minimization problem with the objective function
\begin{equation}\label{Eq: objective function}
    F(\sigma) = \frac{1}{2} \norm{U(\sigma)-V}^{2} + R(\sigma),
\end{equation}
where $V = \left( V_{1}^{(1)}, \dots , V_{L}^{(1)}, \dots , V_{1}^{(K)}, \dots , V_{L}^{(K)} \right)^{T}$ in $\mathbb{R}^{KL}$
represents a vector of the measured voltages on each of the $L$ electrodes for $K$ linearly independent current patterns and $U(\sigma) = \left( U_{1}^{(1)}(\sigma), \dots , U_{L}^{(1)}(\sigma), \dots , U_{1}^{(K)}(\sigma), \dots , U_{L}^{(K)}(\sigma) \right)^{T}$ in $\mathbb{R}^{KL}$ is a vector of the simulated voltages at the $L$ electrodes produced by the same $K$ current patterns using the conductivity $\sigma$. The term $R(\sigma)$ represents a possible regularization term added to the norm of the residuals, also called the data-fidelity term. 

For an initial guess, $\sigma_{0}$, we use the best constant conductivity fit to the data \cite{Jain1997}. Then, \eqref{Eq: objective function} can be rewritten as 
\begin{equation}\label{Eq: objective function 2}
    F(\sigma_{0}+\delta\sigma) = \frac{1}{2} \norm{U(\sigma_{0} + \delta\sigma)-V}^{2} + R(\sigma_{0}+\delta\sigma),
\end{equation}
with the intent to minimize it with respect to $\delta\sigma$.  Solving \eqref{Eq: objective function 2} iteratively leads to the following classic update rule, 
\begin{equation}\label{Eq: classic iteartion}
    \sigma_{k+1} = \sigma_{k} + \delta\sigma_{k},
\end{equation}
given an estimate $\sigma_{k}$. We then iterate until a satisfactory solution for \eqref{Eq: objective function} is found.

\subsubsection{The Levenberg-Marquardt Algorithm}\label{SubSubSec: LM}
%
When there is no explicit regularization included in the objective function \eqref{Eq: objective function}, then $R(\sigma)=0$. The Taylor expansion of \eqref{Eq: objective function 2} to the quadratic term is then given by
%
 \[   F(\sigma+\delta\sigma) = F(\sigma) + F'(\sigma)(\delta\sigma) + \frac{1}{2}F''(\sigma)(\delta\sigma)^{2},\]
%
and a minimum can be found by setting the gradient with respect to $\delta\sigma$ equal to ${0}$. This yields the update
\begin{equation}\label{Eq: Newton method}
    \delta\sigma = -F''(\sigma)^{-1}F'(\sigma),
\end{equation}
where $F'(\sigma)$ and $F''(\sigma)$ are the gradient and Hessian of the objective function $F$. When $R(\sigma)=0$, these are defined as
%
%
\begin{IEEEeqnarray}{lCl}
F'(\sigma) &=& J \left( \sigma \right) ^{T} \left(U \left( \sigma \right) - V \right)\nonumber\\
    F''(\sigma) &=& J \left( \sigma \right) ^{T} J \left( \sigma \right) + \sum_{i} U_{i}''(\sigma) \left(U_{i} \left(\sigma \right) - V_{i} \right), \label{Eq: Fx hessian}
\end{IEEEeqnarray}
%
where $J(\sigma)$ is the Jacobian of the simulated voltages $U(\sigma)$ (e.g., computed by \cite{Kaipio2000, Vauhkonen1997}). In Newton's method, the Hessian is computed exactly according to \eqref{Eq: Fx hessian}. In the Gauss-Newton (GN) method, the second term is ignored due to the costly computation of the second order derivative $U_{i}''(\sigma)$. Alternatively, the Levenberg-Marquardt (LM) algorithm proposes replacing the second term with a scaled identity matrix $\lamLM I$ where $\lamLM \in \mathbb{R}^{+}$, acting as regularizer of the ill-posed problem.  This has the benefit of improving the condition number for the matrix to be inverted in \eqref{Eq: Newton method} which is of particular importance when the Jacobian is rank deficient. The approximate solution to \eqref{Eq: objective function 2} is then
\begin{multline}\label{Eq: LM}
    \delta\sigma_{\mbox{\tiny LM}} = -\left( J \left( \sigma \right) ^{T} J \left( \sigma \right) + \lamLM I \right)^{-1} 
    J \left( \sigma \right) ^{T} \left(U \left( \sigma \right) - V \right).
\end{multline}
We obtain an iterative reconstruction algorithm by using the LM updates \label{Eq: LM} in the update rule \eqref{Eq: classic iteartion} and a suitable stopping criteria. In this work, we will use $\delta\sigma_{\mbox{\tiny LM}}$ for our learned GCNM.

\subsubsection{Regularized Gauss-Newton}\label{SubSubSec: GN}
Alternatively, one can enforce certain priors, such as piecewise constant conductivity reconstructions, that are desirable in EIT, as opposed to smooth reconstructions. This is typically approached via  Total Variation (TV) with $R(\sigma) = \lamTV \sum_{i} \abs{\mathcal{L}_{i} \sigma}$ where $\mathcal{L}$ is a sparse matrix representing the discrete gradient, see \cite{Borsic2010} for details.  One often considers a smoothed approximation of Total Variation regularization using $R(\sigma) = \lamTV \sum_{i} \sqrt{ \left( \mathcal{L}_{i} \sigma \right)^{2} + \gamma }$.  The approximate solution to minimize \eqref{Eq: objective function 2} is then
\begin{multline}\label{Eq: TV}
    \delta\sigma_{\mbox{\tiny TV}} = -( J \left( \sigma \right) ^{T} J \left( \sigma \right) + \lamTV \mathcal{L}^{T}E^{-1} \mathcal{L} )^{-1} 
    \\
    \left( J \left( \sigma \right) ^{T} \left(U \left( \sigma \right) - V \right) + \lamTV \mathcal{L}^{T} E^{-1} \mathcal{L} \sigma \right)
\end{multline}
where $\gamma \in \mathbb{R}^{+}$ is the smoothing parameter that can be varied \cite{Borsic2010}, and $E = \mathrm{diag}\left( \sqrt{ \left( \mathcal{L} \sigma \right)^{2} + \gamma} \right)$ is a diagonal matrix.  We will compare the GCNM recosntructions to TV reconstructions using \eqref{Eq: TV}. 

%
\section{Methods}\label{Sec: Methods}
%
%

\begin{figure}[h!]
\centering
\includegraphics[width=250pt]{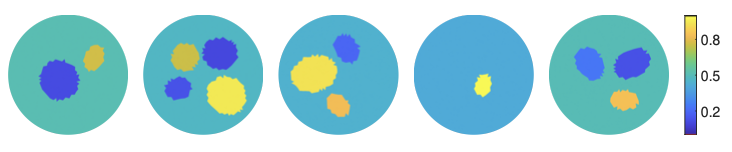}

%
%
%
%
%
%
%
%
%
\caption{\label{Fig: training samples} Examples of "Truths" from the simulated training data.}
\end{figure}

To evaluate the performance of the novel GCNM we looked at a case study for 2D absolute EIT imaging.  We trained a GCNM network using the LM update~\eqref{Eq: LM} using 400 simulated training samples and 100 simulated testing samples.  The simulated samples each contained 1-4 piece-wise constant elliptical inclusions all defined on the same circular mesh (radius 140mm, $L=32$ equally spaced electrodes of width 20mm and height 20mm).  For each simulated conductivity phantom, the forward EIT problem \eqref{Eq: eit main} and \eqref{Eq: complete electrode model} was solved using FEM with approximately 5,000 triangular elements using adjacent current patterns with current amplitude 2mA.  The inverse problem was solved using a FEM mesh with approximately 4,000 elements.  A study of the effect of mesh granularity on reconstruction quality across methods was not explored.   Figure~\ref{Fig: training samples} shows sample simulated phantoms used in the training with conductivity values given in Table~\ref{Tab: inclusion table}.

Prior to solving the inverse problem, noise was added to the simulated voltages of the training data using 
\[v_{p} = v_{p} + \nu \text{mean} \left( \abs{v_{p}} \right) \epsilon_{p}.\]
The parameter $\nu$ controls the level of noise, $v_{p}$ is the vector of voltage measurements on the electrodes for the $p^{th}$ current pattern, and $\epsilon_{p} \in \mathbb{R}^{L}$ is a vector of Gaussian random numbers. Unless otherwise stated, 0.5\% noise ($\nu=0.005$) was used in this study. This corresponds to an SNR of about 51dB.  The KIT4 system has an SNR of approximately 65.52dB~\cite{nissinen2011compensation} which is well above the value used here.

%
\begin{table}[]
\centering
\caption{Inclusion numbers and conductivity values for the samples in each test case are shown below. The conductivity values are presented in S/m and where a range is shown, values were drawn from a uniform distribution within the range.}
\label{Tab: inclusion table}
\begin{tabular}{|l|cccc|}
\hline
                 & Training         & Cases 1-6        & ACT3 & KIT4  \\ \hline
$N_{inclusions}$ & 1-4              & 1-3              & 3  & 2-4     \\ 
$\sigma_{bkg}$   & {[}0.40, 0.43{]} & {[}0.40, 0.43{]} & 0.424 & 0.135 \\ 
$\sigma_{low}$   & {[}0.15, 0.25{]} & {[}0.15, 0.25{]} & 0.24    & 0.061  \\ 
$\sigma_{high}$  & {[}0.65, 0.95{]} & {[}0.65, 0.95{]} & 0.75  & 0.323  \\ \hline
\end{tabular}
\end{table}
%

%
\subsection{Examples Considered}\label{sec:examples}
%

Several test cases will be used to assess the performance of the trained GCNM.  First we consider samples within the distribution of the training data (Case 1). With an eye on the larger question of how well the trained network generalizes to data it has not seen, we focus the remainder of our attention on the following out of distribution cases (Table~\ref{Tab: examples table}).  Cases~2-6 explore samples from simulated data outside the distribution of training data. Namely, Cases~2-3 use a chest shaped domain with perimeter 900mm.  Case~2 assumes correctly known electrode locations but Case~3 assumes that errors are made when placing the electrodes in an experiment. The locations along the boundary of the electrodes are shifted by $N(0\mathrm{mm},1\mathrm{mm})$ when simulating the measured voltages as compared to where they are assumed to be during reconstruction (evenly spaced). The same shift was used for all samples in Case~3. Case~4 explores incorrect domain modeling by reconstructing on an ovular domain (semi-major/minor axes of 170mm and 110mm) while the measured voltages were simulated on the chest shaped domain used in Cases~2-3.  This corresponds to a measurement scenario where the patient shape is not accurately known.  Case~5 introduces sharp corners by using samples with large `L'-shaped targets as opposed to the smaller ovular targets used in training. Case~6 considers samples with varying levels of added noise.  

Lastly, we reconstruct absolute EIT images from experimental data in Case~7 using data collected with the 32 electrode ACT3 system \cite{Cook1994,Isaacson2004} and 16 electrode KIT4 system \cite{hauptmann2017open}, respectively. The conductivity values for all targets are listed in Table~\ref{Tab: inclusion table}.  The archival ACT3 data used trigonometric current patterns with maximum amplitude 0.2mA and frequency 28.8kHz on a tank of radius 150mm, with 32 equally spaced electrodes of width 25mm and saline height 16mm.  The KIT4 data used adjacent current patterns with amplitude 3mA at current frequency 10kHz on 16 approximately equally spaced electrodes.  The circular tank had a radius of 140mm, with electrodes of width 25mm, and had two targets a large resistor (0.067~S/m) and small conductor (0.305~S/m) sitting in a saline bath 0.135~S/m of height 45mm.  The chest shaped tank had perimeter 1020mm,  electrodes of width 20mm, and contained conductive (pink) agar targets of conductivity 0.323~S/m and resistive (white) agar targets of conductivity 0.061~S/m in a saline bath of 0.135~S/m filled to a height of approximately 47mm.  See \cite{Isaacson2004} and \cite{Hamilton_2019} for additional experimental details for the ACT3 and KIT4 data, respectively.

\begin{table}[]
\centering

\caption{Summary of test cases explored. The domain shape, whether or not the mesh is the same as the one used in training, and other information are noted.}
\label{Tab: examples table}
\begin{tabular}{|c|l|l|l|}
\hline
Case &  Mesh & Domain        & Other info:                                            \\ \hline \hline
1            & Old         & Circle  & Consistent with training data.                         \\ \hline
2            & New         & Chest  & Perimeter 900mm.                                      \\ \hline
3            & New         & Chest  & Incorrect domain modeling (electrodes). \\ \hline
\multirow{3}{*}{4}           & \multirow{3}{*}{New}         & \multirow{3}{*}{Oval}  & Incorrect domain modeling.  \\
&&& True: Chest. \\
&&& Recon: Oval (110mm, 170mm). \\ \hline
5            & Old         & Circle  & `L-shaped' targets with sharp corners. \\ \hline
6            & Old         & Circle  & Varying levels of noise (0\%, 1\%, and 2\%).  \\ \hline
\multirow{3}{*}{7}           & New         & Circle  & ACT3 data (r=150mm).  Trig current patterns.   \\ \cline{2-4}
            & New         & Circle & KIT4 data (r=140mm). 
$L=16$ electrodes.  \\ \cline{2-4}
            & New         & Chest & KIT4 data. 
$L=16$ electrodes.  \\ \cline{2-4} \hline 

\end{tabular}
\end{table}

Finally, an important piece of modeling the forward problem is estimating the contact impedance between the electrodes and the surface of the domain. For all of the simulated cases, the contact impedance at each electrode was selected from $N(5\mathrm{\mu\Omega{m}} ,0.5\mathrm{\mu\Omega{m}})$ when computing the measured voltage. Then, for all reconstructions, including the experimental test cases, the contact impedance was assumed to be the mean.  Note that a device/experiment specific tuning of the contact impedance values for each case could yield improved results.  
Here we choose to forgo such tuning to emphasize the generalizability of the GCNM.

\subsection{Comparison methods}
%
%


We compare the learned GCNM results to the classical variational LM and TV approaches as well as a Graph Residual Network (GResNet). For the LM and TV algorithms, the regularization parameters, $\lamLM=10$ for LM and $\lamTV=0.005$ and $\gamma=1\mathrm{e}{-8}$ for TV, were chosen empirically as those which attained a minimum mean squared error (MSE) in conductivity for a subset of the training data. These remained constant except where specified in Case~7. Additionally, a line search was implemented in the classical methods to select the step length in the direction $\delta\sigma$ that minimizes the objective function \eqref{Eq: objective function}. Iteration of the classical methods and the GCNM was stopped with $\sigma_{rec} =\sigma_{k}$ if the objective function failed to decrease on the following 3 iterations. 

%
\subsubsection{A Graph Convolutional Residual Network}\label{SubSec: GResNet}
%
The most computationally expensive step in each iteration of the GCNM for EIT is computing the Jacobian of the forward problem $J(\sigma)$ used in the Newton update $\delta\sigma$. To show the importance of including this model information, the GCNM will also be compared to a GResNet of about the same size. As is common in other residual networks, the GResNet used in this work and shown in Fig.~\ref{Fig: network diagram} used skip connections to add the input and output of each block.  
See \cite{ZhangJiawei2019} for a discussion on skip connections and graph residual network size.

The first iteration of LM algorithm (without the use of a line search) with $\lamLM=0.1$ was used as input to the GResNet with similarly structured blocks as the GCNM. In total, five blocks with the skip connections were used in series to compute the final reconstruction, $\sigma_{rec}$. To be clear, the main difference between the GResNet and the GCNM, is the lack of new information being introduced at each block and can be understood as an ablation study for the model information. The GCNM is a model-based iterative method while the GResNet is a post-processing network. Without the need to compute the updates at each iteration, the GResNet is trained end-to-end.

%

\subsection{Training Details for GCNM and GResNet}\label{SubSec: Training}

The same set of data was used to train both learned methods.  For the GCNM, we computed the updates $\delta\sigma$ following the LM algorithm \eqref{Eq: LM} with $\lamLM=0.1$. Each block was then trained minimising the iterate-wise loss function \eqref{Eq: loss fn} using mini-batches of 10 samples and the Adam optimization method \cite{kingma2015} as it is implemented in PyTorch (version 1.7.1). A learning rate of 0.002 was used. Lastly, training was only stopped when the validation loss failed to decrease for 200 epochs, and the trainable parameters that resulted in the minimum validation loss were saved. No constraints on the trainable parameters were used. 
Training the GCNM with 10 blocks took approximately 12 hours. 
Training of the blocks was done on a NVIDIA Titan V GPU while simulating the forward problem and computing $\delta\sigma$ were performed on the CPU.
Although a system of 10 blocks was trained, the reconstruction was chosen as $\sigma_{rec}=\sigma_{k}$ when the objective function \eqref{Eq: objective function} failed to decrease in the following three iterations
or as the output of the final block $\sigma_{rec}=\sigma_{k_{max}}$ if the first criterion was not met.  

%
\begin{figure}[hb!]
    \centering
    
        {\includegraphics[width=230pt]{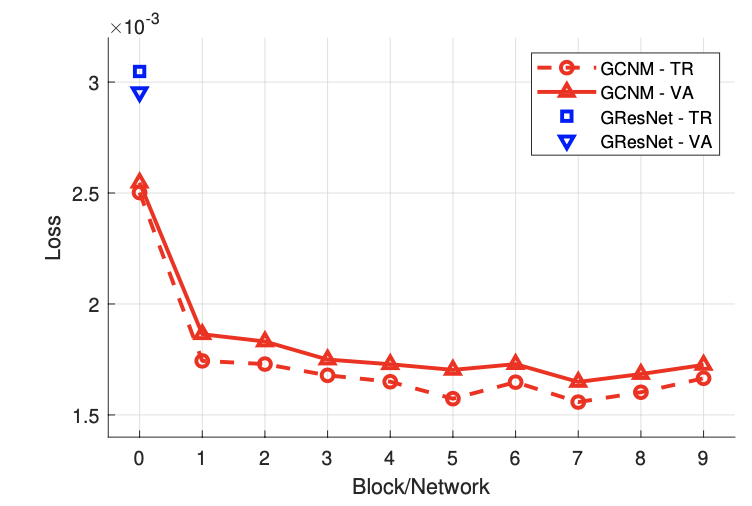}}
        
    \caption{ \label{Fig: training loss} 
        The final loss values (training and validation) of each block of the the GCNM and for the GResNet as a whole. 
        }
\end{figure}

The training for GResNet used the same sized mini-batches, optimizer, and stopping criteria as for training the GCNM networks, and was trained end-to-end in about 6 hours, including the time to solve the forward problem and compute $\delta\sigma_{0}$, on the same hardware. Fig.~\ref{Fig: training loss} displays the training and validation losses at the epoch with the minimum validation loss for each of the 10 blocks in the GCNM and the GResNet as a whole. The losses for the GResNet are shown at Network~0 on the horizontal axis because only one Newton-type update is used, the same as block~0 of the GCNM. The GResNet is not able to achieve a smaller validation loss than the first iteration of GCNM which indicates that a deeper residual network structure with GCN blocks did not perform well and may have difficulties to capture the underlying structures for the EIT reconstruction problem. 

%
%
%
\subsection{Metrics}\label{SubSec: Metrics}
As there is no gold-standard metric for EIT images, we include the following metrics to give information on the quality of the reconstructed conductivity: $\MSEsig$, dynamic range
\[
    DR = \frac{\max(\sigma_{k})-\min(\sigma_{k})}{\max(\sigma_{true})-\min(\sigma_{true})} \times 100\%,\]
%
$\ell_1$ relative conductivity error
\[RE_{\sigma_k}^{\ell_1}= \frac{ \norm{\sigma_{k} - \sigma}_{1} }{ \norm{\sigma}_{1} },\]
as well as the $\ell_2$ relative voltage error 
\begin{equation}\label{Eq: rel error v}
    RE_{V}^{\ell_2}(\sigma_{k})= \frac{ \norm{U(\sigma_{k}) - V}_{2} }{ \norm{V}_{2} } .
\end{equation}
To compute the metrics for the test Cases 1-4, averages over 100 simulated samples will be reported. Case~5 uses an average over only 10 samples due to the lack of deviation in inclusion location and size of the `L-shaped' targets. Case 6 uses an average over 100 samples at each noise level. For the experimental data, the metrics each correspond to a single sample of data.

\begin{figure*}[h!]
\centering
\includegraphics[width=460pt]{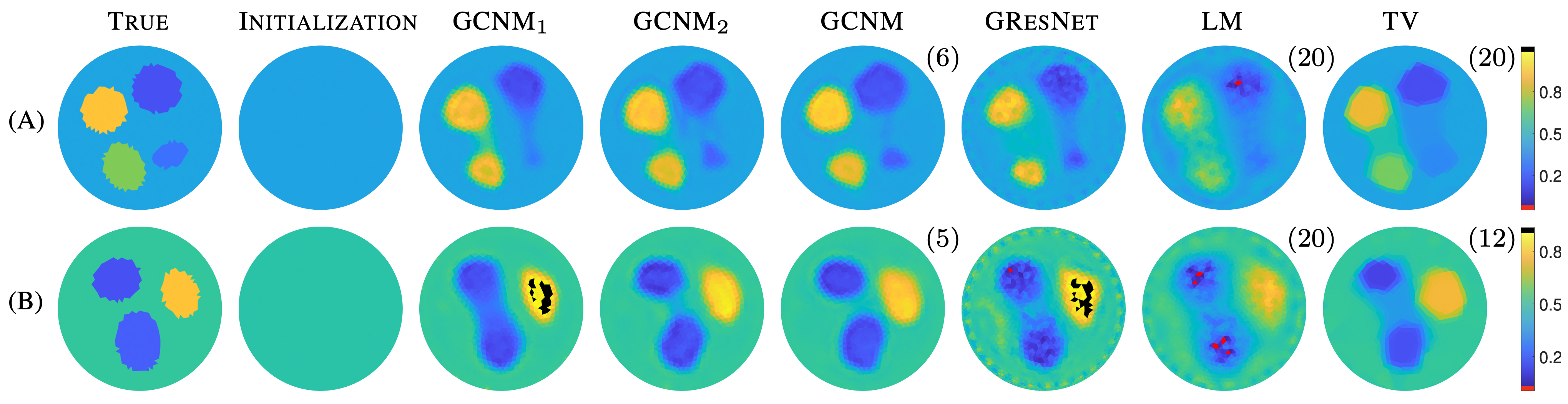}
\caption{\label{fig:results_indist} Case 1: Results for two samples of testing data consistent with, but not included in, the training data. The initialization, first two iterations of the GCNM, and the GCNM reconstruction are compared to the reconstructions from GResNet, LM, and TV. Black and red colors represent going 20\% over or under the color bar range, respectively.  Iteration numbers chosen by the stopping criterion are shown at the top right for each iterative method. 
}
\end{figure*}

%
\section{Results}\label{Sec: Results}
%
%
We now present the results for simulated in and out of distribution data, as well as experimental data that is also out of distribution as the GCNM network and GResNet network were not optimized for the specific EIT devices, target types, contact impedances, current patterns (ACT3), or number of electrodes (KIT4).

%
\subsection{Simulated Results}\label{SubSec: Simulated Results}
Figure~\ref{fig:results_indist} compares the results of the new GCNM against GResNet, LM, and TV for two samples consistent with, but not used in, the training or validation data. 
The initialization as well as the next two iterations of GCNM are shown in addition to the output GCNM image chosen by the stopping criterion.  The number in the top right of each image denotes the iterate chosen by the stopping criterion for each iterative method.  Note that the GCNM outperforms the other methods in visual sharpness as the only method to clearly separate all four and all three targets for Samples A and B, respectively.  Table~\ref{Tab: metrics table} shows that the GCNM required the lowest number of iteration on average and achieved the lowest conductivity errors $MSE_\sigma$ and ${RE}_{\sigma}^{\ell_1}$ of the iterative schemes.  As expected, the classical methods LM and TV both achieved lower relative voltage errors $RE_V^{\ell_2}$ as they minimize over the voltage error but both required more than 15 iterations on average with the LM averaging the full 20 allowed.  It is not unexpected that the GCNM and GResNet do not achieve the lowest relative voltage error as the networks were optimized to reduce the MSE in the conductivity rather than the voltage data. The TV method achieved the best dynamic range closely followed by the GCNM. The GResNet provided significantly improved reconstructions over the initialization in a single application but often overestimated the dynamic range as can be seen in sample B of Figure~\ref{fig:results_indist}.  The output from the GResNet looks similar to the first iteration of the GCNM except with more artefacts, especially near the boundary.

\begin{figure}[h!]
\centering
\includegraphics[width=250pt]{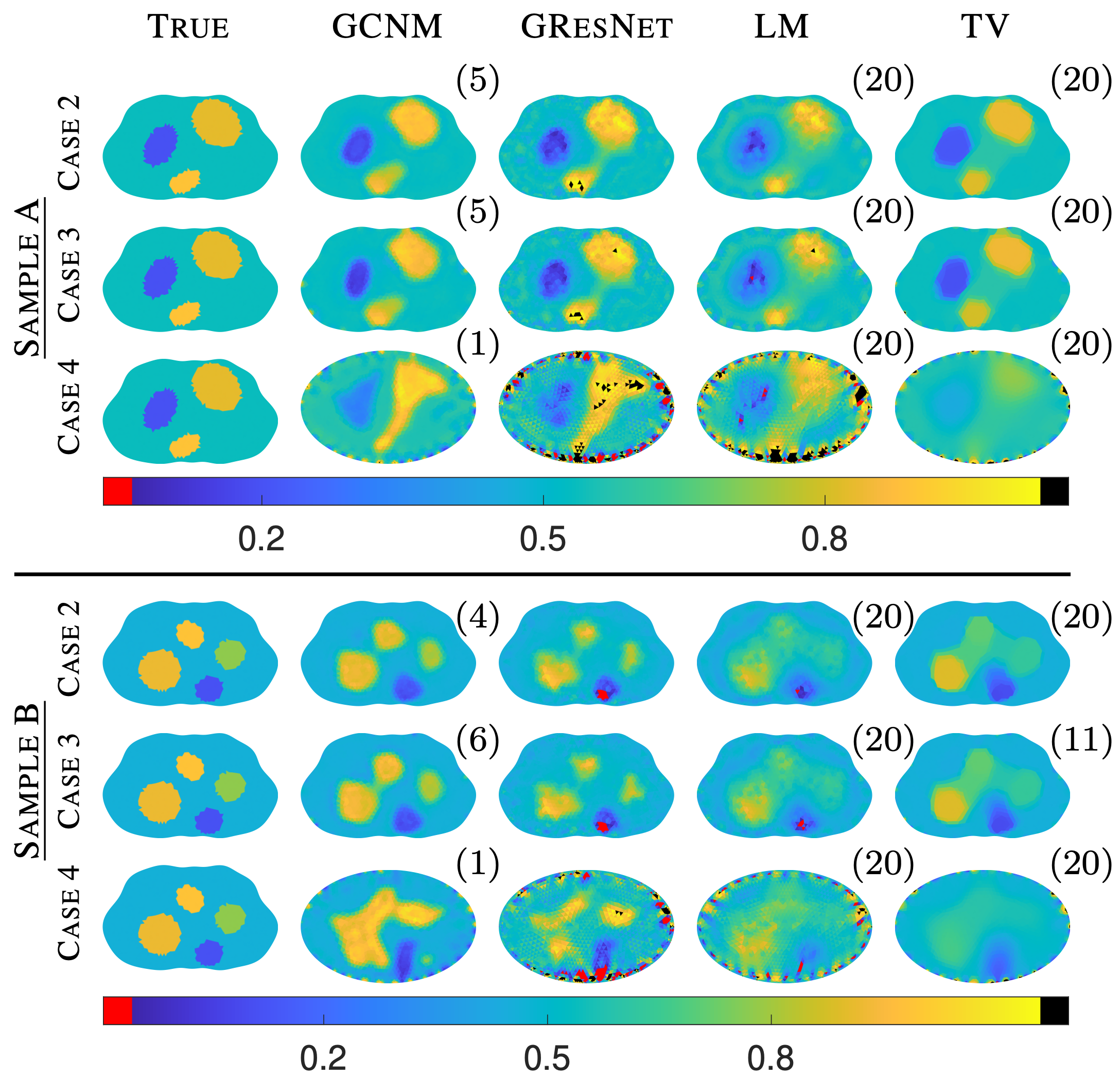}
\caption{\label{fig:recons-outdist-dom} Results for Cases~2-4: exploring a chest shaped domain, and incorrect domain modeling} 
\end{figure}

Next, Figure~\ref{fig:recons-outdist-dom} presents results of the methods on {\it out of distribution} data for Cases~2-4; metrics are presented in Table~\ref{Tab: metrics table}.  Beginning with Case~2, the chest shaped domain, again the GCNM and TV produce the sharpest reconstructions with TV outperforming the other methods in all but the relative voltage error (LM) but requiring an average of 15.8 iterations.  In Sample~B, only the GCNM and GResNet were able to clearly separate all four targets.  By the metrics, the GCNM performed second-best in $\MSEsig$, $\REsig$ and DR in an average of 4.4 iterations.  Recall that the learned networks were optimized for only data coming from a circular domain.  This generalizability stems directly from the translation invariance of convolutions and nicely illustrates the capabilities of the network to handle a mesh that it was not trained on. This holds promise for clinical imaging settings where domain shapes would be inconsistent from patient to patient.  In Case~3, for which the electrode locations are incorrectly modeled, visually the results are similar to that of Case~2 with the GCNM and GResNet again obtaining clearly identifiable separated targets requiring an average of 3.9 iterations and a single application, respectively, and the GCNM obtaining the best dynamic range and second-best $\REsig$.  The traditional LM and TV methods slightly outperformed the learned methods in $\MSEsig$ and $\REv$ again requiring over 16 iterations on average.
 
In Case~4 we reconstruct assuming an ovular domain instead of the true chest shaped domain. Typically, mis-modelling of the domain for absolute EIT results in large artefacts near the boundary of the domain where the mis-modeling has occurred. Here we see major artefacts around the boundary in the GResNet and LM reconstructions.  The GCNM contains boundary artefacts as well, but the effect is less pronounced.   While Figure~\ref{fig:recons-outdist-dom} shows that none of the methods are able to resolve all the targets for these samples, the GCNM appears the most stable followed by GResNet.  The LM method resolves some but not all of the targets but suffers additional artefacts in the center.  The TV reconstructions are dominated by domain mismatch modeling errors at the electrodes resulting in very low contrast, non-separated, reconstructions in the center of the images. Large magnitude errors in pixels at the boundary of the domain led to large dynamic range values that are reflected in Table~\ref{Tab: metrics table noise} with an average of 444.7\%.  By the metrics, the GCNM outperformed all methods in an average of just one iteration, except for the relative voltage error.  As the domain model is such a poor match, we cannot expect the iterates to minimize the objective function. To overcome this, one could consider different stopping criteria for GCNM.  The LM method preformed second-best according to the metrics.  We point out that Cases~3 and 4 are particularly important points of study for EIT where it is unlikely to precisely know the electrode locations and a patient breathing or moving changes their location as well as the domain shape.

\begin{figure}[h!]
\centering
\includegraphics[width=250pt]{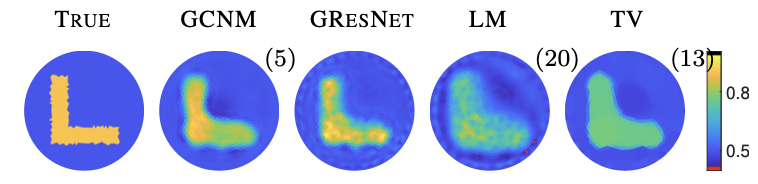}
%
%
%
%
%
%
%
%
%
%
%
%
\caption{\label{Fig: bigL} Results for Case 5: An L-shaped out of distribution inclusion.}
\end{figure}

Figure~\ref{Fig: bigL} explores Case~5 where the target inclusion is an `L', a notoriously challenging target for EIT.  Recall that the training data was only elliptical inclusions and thus this target was quite different than those to which the network was exposed during training.  Nonetheless, each method produces visually recognizable `L' shaped targets.  The GResNet image contains significant fluctuations at the domain boundary, as does LM.  The GCNM contains a low conductivity artefact near the inner corner of the `L', as does the TV image, but still resulted in the sharpest corner reconstruction and best metrics (see Table~\ref{Tab: metrics table}) aside from the voltage error.

\begin{table}[]
\centering
\caption{Error metrics for Cases 1-5. Reported Metrics are averages over 100 samples except for Case 5, which used 10 samples. Bold table entries represent best in case scores.}
\label{Tab: metrics table}
\begin{tabular}{|c|c|ccccc|}
\hline
Case               & Method  & Its  & ${MSE}_{\sigma}$  & ${RE}_{\sigma}^{\ell_1}$ & ${RE}_{V}^{\ell_2}$ & DR(\%)         \\ \hline
\multirow{4}{*}{1} & GCNM    & 4.8  & \textbf{1.71e-03} & \textbf{3.34e-02}        & 9.44e-03            & 113.3          \\
                   & GResNet & 1.0  & 2.95e-03          & 5.75e-02                 & 1.68e-02            & 131.2          \\
                   & LM      & 20.0 & 3.96e-03          & 7.24e-02                 & 2.68e-03            & 124.9          \\
                   & TV      & 15.8 & 2.62e-03          & 4.40e-02                 & \textbf{2.66e-03}   & \textbf{89.8}  \\ \hline
\multirow{4}{*}{2} & GCNM    & 4.4  & 3.21e-03          & 5.21e-02                 & 2.98e-03            & 126.9          \\
                   & GResNet & 1.0  & 5.56e-03          & 7.78e-02                 & 1.19e+00            & 199.4          \\
                   & LM      & 20.0 & 4.25e-03          & 8.08e-02                 & \textbf{2.93e-03}   & 154.8          \\
                   & TV      & 15.8 & \textbf{2.46e-03} & \textbf{4.49e-02}        & 4.78e-03            & \textbf{101.6} \\ \hline
\multirow{4}{*}{3} & GCNM    & 3.9  & 5.22e-03          & 7.45e-02                 & 5.47e-02            & \textbf{171.6} \\
                   & GResNet & 1.0  & 6.29e-03          & 8.82e-02                 & 2.47e-01            & 219.1          \\
                   & LM      & 20.0 & 5.11e-03          & 9.66e-02                 & \textbf{2.79e-03}   & 190.7          \\
                   & TV      & 16.6 & \textbf{4.27e-03} & \textbf{6.72e-02}        & 3.12e-03            & 182.4          \\ \hline
\multirow{4}{*}{4} & GCNM    & 1.0  & \textbf{2.47e-02} & \textbf{2.30e-01}        & 1.68e-01            & \textbf{200.8} \\
                   & GResNet & 1.0  & 6.94e-02          & 3.47e-01                 & 1.38e+01            & 773.0          \\
                   & LM      & 20.0 & 3.55e-02          & 2.80e-01                 & \textbf{8.77e-03}   & 361.5          \\
                   & TV      & 20.0 & 5.98e-02          & 3.12e-01                 & 8.01e-02            & 444.7          \\ \hline
\multirow{4}{*}{5} & GCNM    & 5.0  & \textbf{1.59e-03} & \textbf{3.06e-02}        & 1.26e-02            & 134.5          \\
                   & GResNet & 1.0  & 2.08e-03          & 4.78e-02                 & 1.44e-02            & 169.9          \\
                   & LM      & 20.0 & 6.87e-03          & 1.14e-01                 & \textbf{2.42e-03}   & 292.5          \\
                   & TV      & 16.0 & 2.73e-03          & 4.87e-02                 & 2.55e-03            & \textbf{102.2} \\ \hline
\end{tabular}
\end{table}

\begin{figure}[h!]
\centering

\includegraphics[width=240pt]{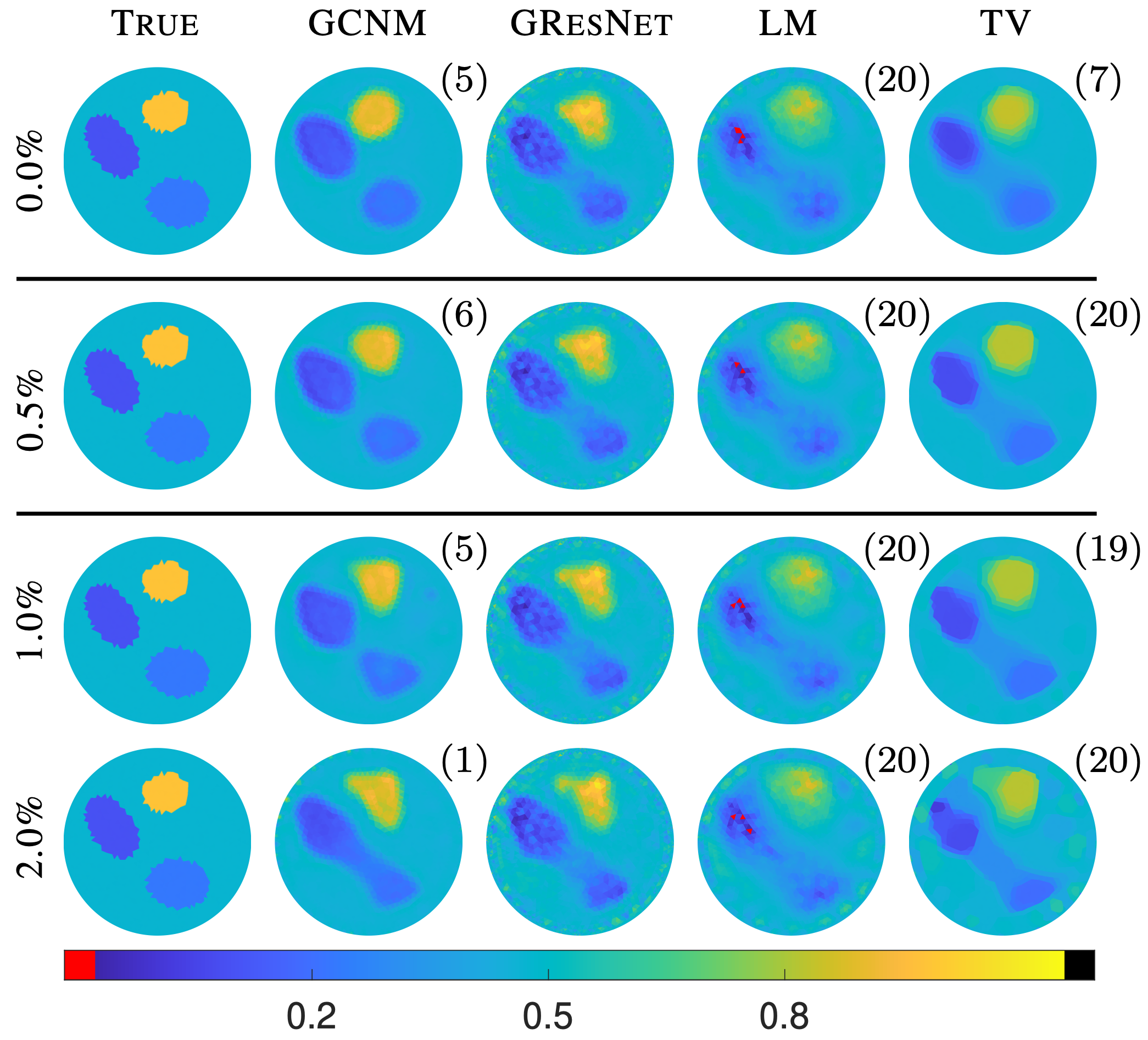}
\caption{\label{fig:recons_noisy} Results for Case 6: Varying the noise level.  Note that samples with 0.5\% noise were used during training.}
\end{figure}

%
%
Figure~\ref{fig:recons_noisy} compares conductivity reconstructions for varying levels of noise in the simulated voltage data corresponding to Case~6.  Recall that the network was trained with $0.5\%$ noise (51dB SNR).  While all the methods performed admirably on this test, the  GCNM was the only method able to separate the three targets clearly, up through 1\% noise (45dB SNR), whereas the other methods blurred the two resistive objects.    

\begin{figure}[h!]
\centering
\includegraphics[width=250pt]{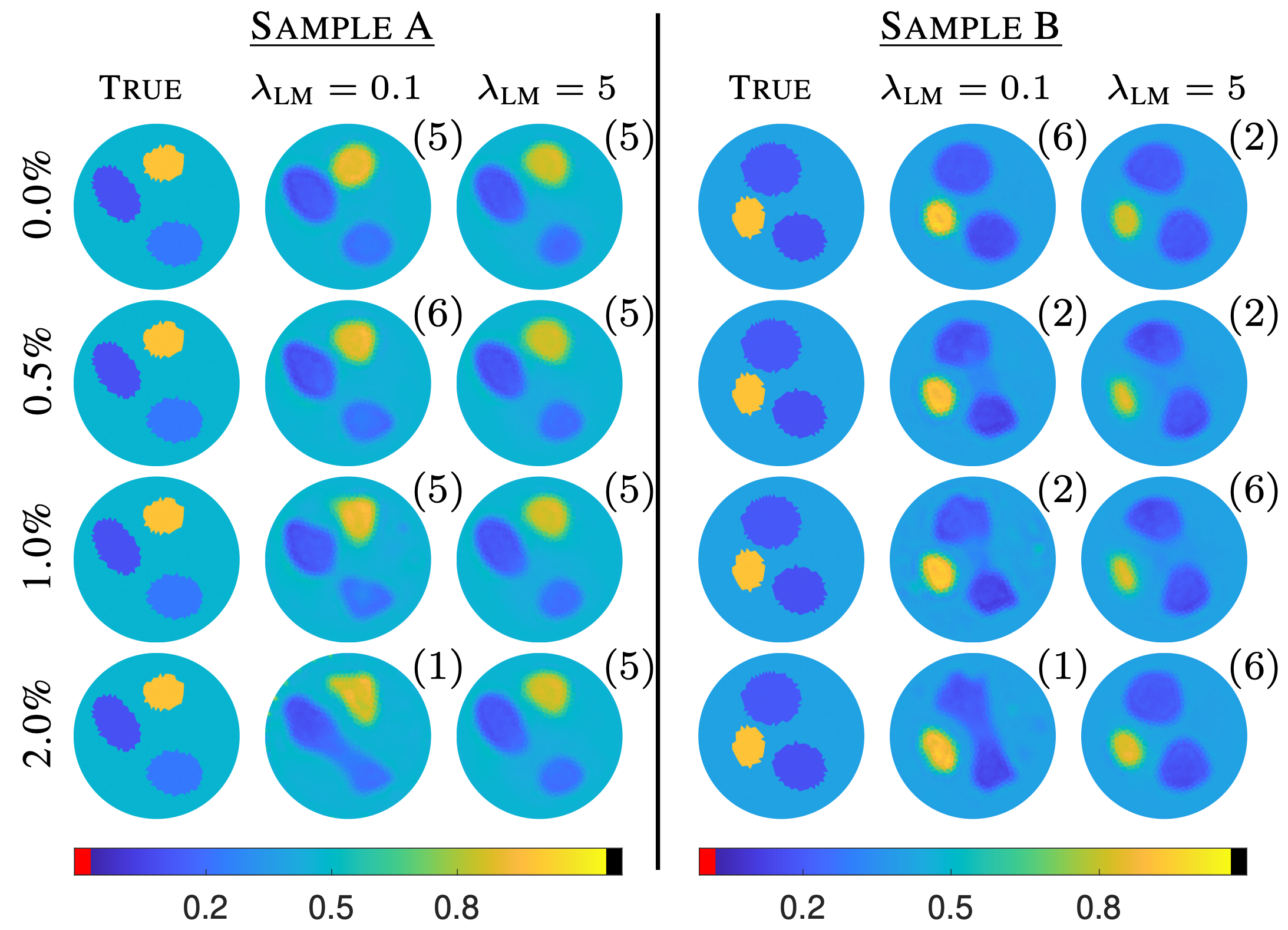}
\caption{\label{Fig: noisy_changing_lambda} Case 6:  GCNM reconstructions for two representative samples with varying levels of added noise shown with two different values for the regularization parameter $\lamLM$ in \eqref{Eq: LM}. Note that the parameter was only changed when testing; all training was done with $\lambda=0.1$.}
\end{figure}

It should be noted that 2\% noise (39dB SNR) is an extreme amount of noise for EIT reconstruction with systems capable of SNR of 65dB, comparable to $0.1\%$ noise \cite{nissinen2011compensation}.  Nevertheless, the GCNM images are not corrupted by errors, and as shown in Figure~\ref{Fig: noisy_changing_lambda}, adjusting the reconstruction parameter $\lamLM$ in the update $\delta\sigma$ in \eqref{Eq: LM} resulted in even fewer artefacts and more clearly defined targets. 

Table~\ref{Tab: metrics table noise} demonstrates that the GCNM and GCNM2 methods outperformed the other methods for all non-voltage metrics except the dynamic range in the 1\% noise case where TV was superior and GCNM2 was second.   The GResNet, in just one iteration, performed well across noise levels while the GCNM and GCNM2 used approximately 5 iterations and TV averaged more than 15 (for $\nu>0$).  The LM method ranked in the top two best for $\REv$ but fell behind in the remaining metrics, as well as visual sharpness requiring, on average, the full 20 iterations.

Note that above 1.0\% noise, using a larger regularization parameter $\lamLM=5$ resulted in reconstructions with lower MSE and relative $\ell_1$ error with GCNM2 vs GCNM. This is a key feature of the proposed GCNM as a model-based approach, that even after training a network with a fixed regularization parameter one can adjust and control the performance if noise levels and distributions change, offering greater flexibility and control. 

\begin{table}[]
\centering
\caption{Metrics for varying the noise level and adjusting $\lamLM$. The method "GCNM" used $\lambda=0.1$ for computing $\delta\sigma$ (same as training) while "GCNM2" used $\lambda=5$.}
\label{Tab: metrics table noise}
\begin{tabular}{|c|c|ccccc|}
\hline
Case               & Method  & Its  & ${MSE}_{\sigma}$  & ${RE}_{\sigma}^{\ell_1}$ & ${RE}_{V}^{\ell_2}$ & DR(\%)         \\ \hline
\multirow{5}{*}{0.0\%} & GCNM    & 4.7  & \textbf{1.56e-03} & \textbf{3.04e-02} & 9.47e-03          & 112.7         \\
                       & GCNM2   & 4.7  & 3.06e-03          & 4.20e-02          & 9.18e-03          & \textbf{94.7} \\
                       & GResNet & 1.0  & 2.88e-03          & 5.60e-02          & 1.62e-02          & 128.4         \\
                       & LM      & 20.0 & 3.90e-03          & 7.00e-02          & 6.09e-04          & 122.3         \\
                       & TV      & 11.1 & 2.53e-03          & 4.20e-02          & \textbf{3.47e-04} & 88.2          \\ \hline
\multirow{5}{*}{0.5\%} & GCNM    & 4.8  & \textbf{1.72e-03} & \textbf{3.34e-02} & 9.44e-03          & 113.3         \\
                       & GCNM2   & 4.7  & 3.06e-03          & 4.21e-02          & 9.62e-03          & \textbf{95.0} \\
                       & GResNet & 1.0  & 2.96e-03          & 5.76e-02          & 1.68e-02          & 131.2         \\
                       & LM      & 20.0 & 3.96e-03          & 7.24e-02          & 2.68e-03          & 124.9         \\
                       & TV      & 15.8 & 2.62e-03          & 4.41e-02          & \textbf{2.66e-03} & 89.8          \\ \hline
\multirow{5}{*}{1.0\%} & GCNM    & 4.2  & \textbf{2.22e-03} & \textbf{4.00e-02} & 1.17e-02          & 117.7         \\
                       & GCNM2   & 4.8  & 3.05e-03          & 4.24e-02          & 1.09e-02          & 95.1          \\
                       & GResNet & 1.0  & 3.21e-03          & 6.21e-02          & 1.83e-02          & 135.0         \\
                       & LM      & 20.0 & 4.15e-03          & 7.79e-02          & \textbf{5.25e-03} & 130.5         \\
                       & TV      & 18.1 & 2.99e-03          & 5.41e-02          & 5.26e-03          & \textbf{98.3} \\ \hline
\multirow{5}{*}{2.0\%} & GCNM    & 1.6  & 3.22e-03          & 4.90e-02          & 2.43e-02          & 118.2         \\
                       & GCNM2   & 5.0  & \textbf{3.08e-03} & \textbf{4.30e-02} & 1.45e-02          & \textbf{96.0} \\
                       & GResNet & 1.0  & 4.22e-03          & 7.69e-02          & 2.82e-02          & 154.4         \\
                       & LM      & 20.0 & 4.90e-03          & 9.27e-02          & \textbf{1.04e-02} & 147.4         \\
                       & TV      & 19.7 & 4.84e-03          & 8.92e-02          & \textbf{1.04e-02} & 135.3         \\ \hline
\end{tabular}
\end{table}

%
\subsection{Experimental Results}\label{SubSec: Experimental Results}
\begin{figure}[h!]
\centering
\includegraphics[width=250pt]{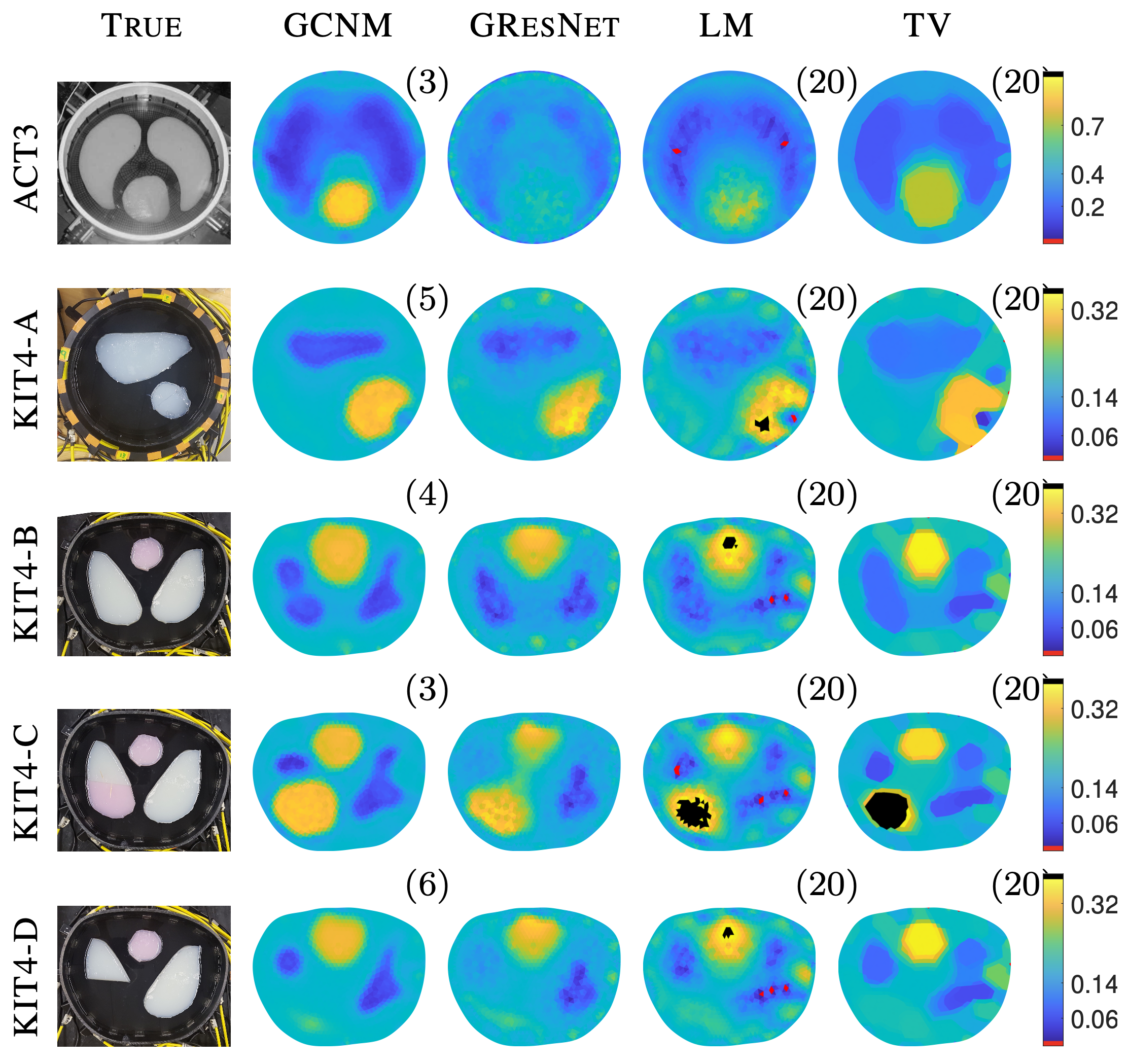}
\caption{\label{fig:recons_experimental} Results for Case 7: Experimental Data.  Column~1 contains the experimental setups. The ACT3 sample is shown with a colorbar range of [0, 1] while the KIT4 samples are shown with colorbar ranges of +/- 20\% of the true range.
}
\end{figure}

We considered five data sets from two different EIT machines using three tanks, the 32-electrode ACT3 system from Rensselaer Polytechnic Institute and the 16-electrode KIT4 system from the University of Eastern Finland.  Results are shown in Figure~\ref{fig:recons_experimental}.  Recall that for all reconstruction methods these are {\it absolute} EIT images, not difference images, and thus they are notoriously challenging to obtain without a careful tuning of the forward model to the specific EIT machine hardware.  Here, the forward model was not tuned to the respective EIT machines and the same contact impedance was used for all electrodes, the mean of those used in the training. 

As the noise distributions and parameter ranges are different for the experimental data when compared to the training data, it is natural that the regularization parameters would need to be tuned for all methods. For the ACT3 reconstructions, $\lamLM=10$ for the GCNM and GResNet, $\lamLM=100$ for the LM algorithm, and $\lamTV=0.02$ for the TV method.  All of the KIT4 samples used $\lamLM=100$ for the GCNM and GResNet, $\lamLM=50$ for the LM algorithm, and $\lamTV=0.02$ for the TV method. The TV methods for both experimental setups used $\gamma=1\mathrm{e}{-8}$, the same as was used for the simulated data.  
Here it is important to emphasize again that the networks of the learned methods were not retrained with these new regularization parameters. The ability to adjust the regularization parameters used in the update terms after training adds to the flexibility of the GCNM and GResNet.  To compute the $\MSEsig$, $\REsig$ and $\REv$ metrics, a simulated truth image was created using the photographs from the experiments and the measured conductivity values reported for the agar targets and saline.  Results are presented in Table~\ref{tab: experimental samples}.  Note that the LM and TV methods required all 20 iterations whereas the GCNM ranged from 3 to 6 iterations across samples.

For the ACT3 data, each reconstruction method recovered the heart and lungs' shapes but failed to completely separate the large resistive lungs. The GCNM and TV methods produced the most distinct target boundaries and the lowest $\MSEsig$ but overestimated the dynamic range.  The GCNM obtained the lowest relative $\ell_1$ error $\REsig$ and required 3 iterations.  
The GResNet had the least distinct targets visually and significantly underestimating the conductivity of the heart,  but the best DR. 
Similarly to the simulated data cases, the LM and TV methods outperformed the learned methods in relative voltage error $RE_V$ requiring 20 iterations each.

The background conductivity in the KIT4 data was 0.135~S/m whereas the training, testing, and ACT3 data was in the range [0.40, 0.43]~S/m.  For the learned methods, to bring the input data into scale for the trained networks, the network inputs $\sigma_k$ and $\delta\sigma_k$ were scaled up by a factor of 3 and the outputs $\sigma_{k+1}$ were scaled back down by a factor of 3; see Figure~\ref{fig:recons_experimental}.   

Beginning with the circular tank, both targets were well located by the all methods for the KIT4-A data.  The GResNet and GCNM obtained the best $\MSEsig$, $\REsig$, and DR.  There are clear artefacts present in bottom right corner of the tank for the LM and TV images.  More broadly, for all of the KIT4 reconstructions, all methods contain an artefact that region suggesting a poor matching of the forward model and/or contact impedance in that region.  A more device-specific tuning of the forward solver to the experimental KIT4 device and contact impedances may improve the results.   

Next, for the KIT4-B data on the chest domain, the two resistive lungs and conductive heart are visible for all methods, each containing slightly different visual artefacts.  The GResNet obtained the lowest $\MSEsig$ and $\REsig$ and the GCNM the best dynamic range. The LM reconstruction contained the least distinct targets.  Sample KIT4-C contained a conductive portion in the bottom left (viewer's) lung.  The LM and TV method significantly overestimated that portion as can be seen by the black pixels in Figure~\ref{fig:recons_experimental} but as usual achieved the lowest relative voltage error $\REv$.  The GResNet achieved the best $\MSEsig$ and $\REsig$ but the top of the viewer's left lung not clearly visible or defined.  The GCNM overestimated the size of the conductive portion of the lung but obtained the second best $\MSEsig$, $\REsig$ and the best DR.   Lastly, reconstructions for sample KIT4-D from the GCNM, LM, and TV methods all show the correct targets with the GResNet underestimating the contrast of the viewer's left lung.  The GCNM and GResNet outperformed the classic methods across metrics except for the relative voltage error.  Summarizing across the KIT4 samples, the GCNM obtained the best DR for all KIT4 samples, and routinely performed in the top two for both $\MSEsig$ and $\REsig$, while only requiring 3-6 iterations compared to the 20 required by the classic methods.

\begin{table}[]
\caption{Table of metrics for each of the experimental samples from ACT3 and KIT4.} 
\label{tab: experimental samples}
\begin{tabular}{|c|c|ccccc|}
\hline
Case  & Method  &  Its  &  ${MSE}_{\sigma}$  &  ${RE}_{\sigma}^{\ell_1}$  &  ${RE}_{V}^{\ell_2}$  &  DR(\%)          \\ \hline
\multirow{4}{*}{\rotatebox{90}{ACT3}}   & GCNM    & 3   & \textbf{1.11e-02} & \textbf{1.76e-01} & 6.58e-02          & 167.7          \\
                        & GResNet & 1   & 1.77e-02          & 2.60e-01          & 1.96e-01          & \textbf{104.8} \\
                        & LM      & 20  & 1.37e-02          & 2.56e-01          & 4.84e-03          & 166.6          \\
                        & TV      & 20  & \textbf{1.11e-02} & 2.11e-01          & \textbf{4.34e-03} & 112            \\ \hline
\multirow{4}{*}{\rotatebox{90}{KIT4-A}} & GCNM    & 5   & 2.40e-03          & 3.07e-01          & 3.38e-02          & \textbf{110.8} \\
                        & GResNet & 1   & \textbf{2.16e-03} & \textbf{2.95e-01} & 3.46e-02          & 124            \\
                        & LM      & 20  & 2.87e-03          & 3.27e-01          & \textbf{3.13e-02} & 180.6          \\
                        & TV      & 20  & 4.28e-03          & 3.93e-01          & \textbf{3.13e-02} & 223.8          \\ \hline
\multirow{4}{*}{\rotatebox{90}{KIT4-B}} & GCNM    & 4   & 3.53e-03          & 3.96e-01          & 3.10e-02          & \textbf{104.6} \\
                        & GResNet & 1   & \textbf{2.91e-03} & \textbf{3.56e-01} & 3.91e-02          & 120.5          \\
                        & LM      & 20  & 3.35e-03          & 3.91e-01          & \textbf{2.66e-02} & 166.1          \\
                        & TV      & 20  & 3.95e-03          & 4.26e-01          & \textbf{2.66e-02} & 156.1          \\ \hline
\multirow{4}{*}{\rotatebox{90}{KIT4-C}} & GCNM    & 3   & 3.11e-03          & 3.74e-01          & 3.27e-02          & \textbf{104.8} \\
                        & GResNet & 1   & \textbf{2.91e-03} & \textbf{3.55e-01} & 3.75e-02          & 124.9          \\
                        & LM      & 20  & 3.34e-03          & 3.84e-01          & \textbf{2.72e-02} & 157.6          \\
                        & TV      & 20  & 3.76e-03          & 4.10e-01          & \textbf{2.72e-02} & 150.7          \\ \hline
\multirow{4}{*}{\rotatebox{90}{KIT4-D}} & GCNM    & 6   & 3.11e-03          & \textbf{2.93e-01} & 2.94e-02          & {\bf 117.5}          \\
                        & GResNet & 1   & \textbf{2.91e-03} & 3.07e-01          & 3.21e-02          & 126.1          \\
                        & LM      & 20  & 3.23e-03          & 3.23e-01          & \textbf{2.58e-02} & 223.1          \\
                        & TV      & 20  & 3.41e-03          & 3.27e-01          & \textbf{2.58e-02} & 169.9          \\ \hline
\end{tabular}
\end{table}


%
%
%
%
%

\vspace{-2em}
\subsection{Further Discussion}\label{SSec: Discussion}
%
%
%
All computations were performed on the hardware described in Section~\ref{SubSec: Training}.  Computational costs per sample were as follows: 2.4 seconds per iteration of GCNM, about 12.3 seconds per iteration of LM and TV, and 4.6 seconds per reconstruction using GResNet.  Each forward solve took approximately 1.1 seconds and each Jacobian computation approximately 1.3 seconds. We emphasize that forward solver and the individual algorithms were not optimized for speed in this study. Nonetheless, the GCNM provided a significant speedup in reconstruction time by requiring approximately five iterations (at 2.4 sec/iter) instead of more than 15 for TV and the full 20 for LM, and without the need to perform a line search at each iteration.  Therefore, the approximate computation time per image for GCNM was 12 seconds the same approximate cost of a single iteration of LM and TV, as implemented in this study.  The GCNM reliably produced images with clearly identifiable targets, and the best or second best metrics across methods for the simulated and experimental data considered in this work, with only a small, general, training set used.  The flexibility of the approach to later allow the user to adjust the regularization parameter(s) of the Newton-type method used (LM here) without retraining the network adds generalizability.  Of particular importance to absolute EIT imaging were the incorrect domain modeling cases (3 and 4).  The apparent robustness of GCNM to such errors holds promise for clinical absolute EIT imaging.  
%
%
%
%
%
%
\section{Conclusion}\label{Sec: Conclusion}
%
%
%

We successfully introduced a novel approach to combine optimization-based solution methods with deep learning directly on nonuniform problem-specific solution meshes. The proposed GCNM provides a simple, yet highly flexible, network architecture that combines the current iterate $\sigma_k$ and its Newton-type update $\delta\sigma_k$ to produce a new improved estimate. This enables the network to leverage the information from both inputs, learn a task-specific prior from training data, as well as improved robustness with respect to noise and potential model mismatch.  As the domain modeling is done outside the network, a network can be trained on a simplified domain such as a circle, and learned weights can still be used later on different domain shapes, without the need for sub-optimal embeddings to rectangular domains. This holds promise for several medical imaging applications where training data could be simulated on a fixed/average domain and then applied to patient-specific models.  %

While we used a simple prototypical network architecture to demonstrate the effectiveness of the model-based approach, extensions to different architectures can be considered as well as additional information supplied to the network. Finally, the approach extends naturally into 3D with no major change, as the graph structure is not limited by spatial dimensions. 
In terms of the specific application, EIT is a highly ill-posed nonlinear inverse problem and as such represents an especially challenging case study. The presented results demonstrate a promising improvement for absolute EIT imaging given that we did not tune the forward solver to the experimental data taken from the two separate machines and three tanks. Additionally, we trained only on general ellipse inclusions and were able reconstruct targets of different shapes. This can be attributed to the additional information supplied by the Newton-type update. We expect that the presented approach extends directly to, and bears great promise for, other tomographic reconstruction problems, where data and image are closely tied to problem specific finite element meshes, instead of pixel/voxel grids.

%
%
%

\section*{Acknowledgement} 
We gratefully acknowledge the support of NVIDIA Corporation with the donation of the Titan Xp GPU used for this research.  We also thank the EIT groups at RPI and UEF for sharing the respective experimental data sets, in particular Ville Kolehmainen of UEF for his helpful discussions on the TV method.

\bibliographystyle{IEEEtran}
\small
\bibliography{IEEEabrv,bib_clean}

\end{document}